# Controllable Antiferromagnetic-Ferromagnetic phase transition in monolayer $MnPSe_3$ via atomic adsorption of Li, O, and F

Dong Liu, Sike Zeng, Ji-Hai Liao and Yu-Jun Zhao*

Department of Physics, South China University of Technology, Guangzhou 510640, China

*Corresponding author: zhaoyj@scut.edu.cn

**ABSTRACT**: The engineering of magnetic order and electronic states in two-dimensional (2D) materials is pivotal for advanced spintronic technologies. Despite their potential, the scarcity of intrinsic 2D ferromagnets remains a critical challenge. Here, we employ density functional theory with Hubbard-*U* corrections to systematically investigate adsorbate-driven magnetic transitions in monolayer $MnPSe_3$. While pristine $MnPSe_3$ exhibits antiferromagnetic (AFM) ordering with a semiconducting gap, Li, O, and F adatom adsorption induces an AFM-to-ferromagnetic (AFM-FM) phase transition across large coverage ranges (25%–100%). Our calculations reveal enhanced thermodynamic stability at elevated coverages, with full-coverage configurations ($Li_{0.5}MnPSe_3$, $MnPSe_3O_{0.5}$, $MnPSe_3F_{0.5}$) favored energetically. Hybrid functional (HSE06) calculations show that F adsorption drives a semiconductor-to-half-metal transition, whereas Li and O adsorption preserves semiconductivity. Moreover, Li adsorption induces a valley splitting of 20.3 meV at the K and K' points in the band structure of the monolayer $MnPSe_3$. Magnetic anisotropy analysis reveals adsorbate-dependent easy-axis reorientation: Li (electron doping) switches the easy-axis from in-plane to out-of-plane, while O/F (hole doping) stabilizes in-plane easy-axis, consistent with carrier-density-modulation simulations. Crucially, carrier doping results indicate that once the electron doping concentration reaches $8.3 \times 10^{13}$ $cm^{-2}$, the magnetic easy axis of monolayer $MnPSe_3$ transitions from in-plane to out-of-plane. This work establishes atomic adsorption as a robust strategy for tailoring 2D magnetism, resolving discrepancies through rigorous treatment of exchange-correlation effects and configurational diversity.

## I. INTRODUCTION

Two-dimensional (2D) van der Waals magnetic materials have garnered significant attention in recent years, and the manipulation of their magnetism is essential for understanding their physical properties and realizing their practical applications [1-3]. As stated by the Mermin-Wagner theorem, long-range magnetic order could not exist in 2D isotropic systems at finite temperature due to thermal fluctuation [4]. Therefore, the magnetic anisotropy induced by spin-orbit coupling is essential for maintaining long-range magnetic order in 2D materials. The successful synthesis of 2D ferromagnetic (FM) materials such as $Cr_2Ge_2Te_6$ [5], $CrI_3$ [6] and $Fe_3GeTe_2$ [7] in experiments have greatly promoted the researches of spintronics. Unfortunately, most of the discovered 2D intrinsic FM materials are metallic and unstable under ambient conditions, which has impeded the further advancement of spintronics [5,6,8].

It is well established that half-metals with one conductive channel and one insulating spin channel, as well as novel van der Waals (vdW) FM semiconductors have been considered as ideal materials for spintronics applications [9,10]. To this end, the exploration of novel vdW FM half-metals and semiconductors are desired for the realization of 2D spintronic devices [11,12]. The typical approaches to manipulate magnetism in 2D materials is through electric field, mechanical stain, chemical modification, defect doping, atom adsorption and magnetic field [13-18]. Among the aforementioned methods for manipulating their magnetic properties, surface functionalization via physical and chemical adsorption is an effective approach, owing to the large specific surface area of 2D materials [19].

Bulk $MnPSe_3$ is a van der Waals magnetic material, previous neutron scattering experiments revealed a collinear antiferromagnetic (AFM) order for its layered system [20]. The interlayer spacing between the magnetic atomic layers is about 6.7 Å, making it an ideal candidate for 2D magnetism. Few-layer $MnPSe_3$ has already been synthesized experimentally through mechanical exfoliation [21]. Monolayer $MnPSe_3$, with a Néel AFM ground state, has been investigated for potential AFM-FM transitions via various methods. For instance, AFM-FM phase transitions can be induced through gate-controlled electron or hole doping [10,18]. Recent studies have demonstrated that by applying a vertical electric field, substituting Se atoms with S to construct a Janus structure, or forming heterostructures with other 2D materials, altermagnetism (AM) can be induced in the monolayer $MnPSe_3$ [22]. Moreover, as demonstrated in recent studies [23,24], spontaneous valley polarization has been successfully achieved in $MnPSe_3$-$WSe_2$ and $MnPSe_3$-$HfN_2$ heterostructures. Therefore, it is considered a highly promising 2D material for spintronic and valleytronic devices. However, as an effective regulation approach, the effect of chemical adsorption on this system have been rarely studied.

In this work, we have achieved a controlled AFM-FM transition in monolayer $MnPSe_3$ through chemical adsorption of Li, O, and F atoms. The thermodynamic stability of Li, O, and F adatoms on 2D $MnPSe_3$ across varying coverage regimes is systematically demonstrated. Notably, full-coverage configurations ($Li_{0.5}MnPSe_3$, $MnPSe_3O_{0.5}$ and $MnPSe_3F_{0.5}$) exhibit superior thermodynamic stability, prompting detailed studies of their magnetic and electronic properties. Heyd-Scuseria-Ernzerhof hybrid functional (HSE06) band structure calculations reveal that while Li and O adsorption with coverage of 100% induces AFM-FM transitions, the semiconducting nature persists. In contrast, F adsorption with coverage of 100% converts $MnPSe_3$ from a AFM semiconductor to a FM half-metal. Notably, Li adsorption at full coverage generates a valley polarization of ~20.3 meV at the K/K′ points of the $MnPSe_3$ band structure. Carrier doping simulations further reveal that magnetic easy-axis reorientation from in-plane to out-of-plane occurs at electron doping concentrations exceeding $8.3 \times 10^{13}$ $cm^{-2}$. Essentially, when incorporating diverse AFM spin configurations (stripy, zigzag), the critical electron/hole concentrations required for AFM-FM transitions substantially exceed earlier theoretical predictions. These findings not only resolve discrepancies in the literature but also establish atomic adsorption as a robust strategy for tailoring spin-ordering and electronic properties in 2D AFM semiconductors, advancing their potential for spintronic and valleytronic applications.

## II. COMPUTATIONAL DETAILS

Our calculations are carried out within the Perdew-Burke-Ernzerhof generalized gradient approximation (GGA) [25] implemented in Vienna ab initio simulation package (VASP) [26], along with HSE06 for the band structure [27]. An energy cutoff of 500 eV for plane waves is adopted. The slab model of system is constructed using 20 Å vacuum along the [001] direction (i.e., *z* direction), to prevent periodic image interactions between the neighboring layers. Monkhorst k-mesh grids of $9 \times 9 \times 1$, $6 \times 9 \times 1$, $6 \times 6 \times 1$ and $4 \times 4 \times 1$ are used for Brillouin-zone sampling of $1 \times 1 \times 1$, $2 \times 1 \times 1$, $\sqrt{3} \times \sqrt{3} \times 1$ and $2 \times 2 \times 1$ $MnPSe_3$, respectively. For geometry optimization, all of the internal coordinates are relaxed until the Hellmann-Feynman force is less than 0.01 $eV \cdot Å^{-1}$ for each ion, and total energy convergence is within $10^{-6}$ eV. Spin-orbit coupling (SOC) is applied in the calculations of magnetic anisotropic energy (MAE) [28,29]. When calculating magnetic properties, the effective Hubbard U correction applied on the 3*d* electrons of Mn atoms is 4 eV follow References [30-32]. Our Néel temperature simulations of pristine monolayer $MnPSe_3$ are carried out using the VAMPIRE atomistic simulation software package [33,34].

## III. RESULTS AND DISCUSSION

### A. Geometric, electronic structure, and magnetic ground state

The top and side views of the crystal structure of the monolayer $MnPSe_3$ are shown in Fig. 1a, with the area enclosed by the red dashed lines representing the unit cell. Each unit cell contains two Mn atoms, six Se atoms, and two P atoms. The lattice parameters of the optimized structure are $a = b = 6.457$ Å, in agreement with recent reported value of $a = b = 6.387$ Å [16]. Clearly, the 2D $MnPSe_3$ layer consists of two $Mn^{2+}$ ions arranged in a hexagonal honeycomb lattice, along with a $[P_2Se_6]^{4-}$ bipyramid, which is formed by a P-P dimer connected to two selenium trimers that vertically pass through the center of each honeycomb plane. To determine the magnetic ground state of 2D $MnPSe_3$, four phases with different magnetic orders—FM, zigzag antiferromagnetic (zigzag-AFM), Néel antiferromagnetic (Néel-AFM), and stripy antiferromagnetic (stripy-AFM)—are constructed using a 2 × 2 × 1 supercell, as shown in Fig. 1b. According to our calculations, the Néel-AFM state exhibits the lowest energy among the four phases, while the energy of the FM state is approximately 21.9 meV/formula unit (f.u.) higher, consistent with previous reports [35-37]. The local magnetic moment of Mn atom is about 4.6 $\mu_B$ where Mn possesses about +2e charge with a high spin state.

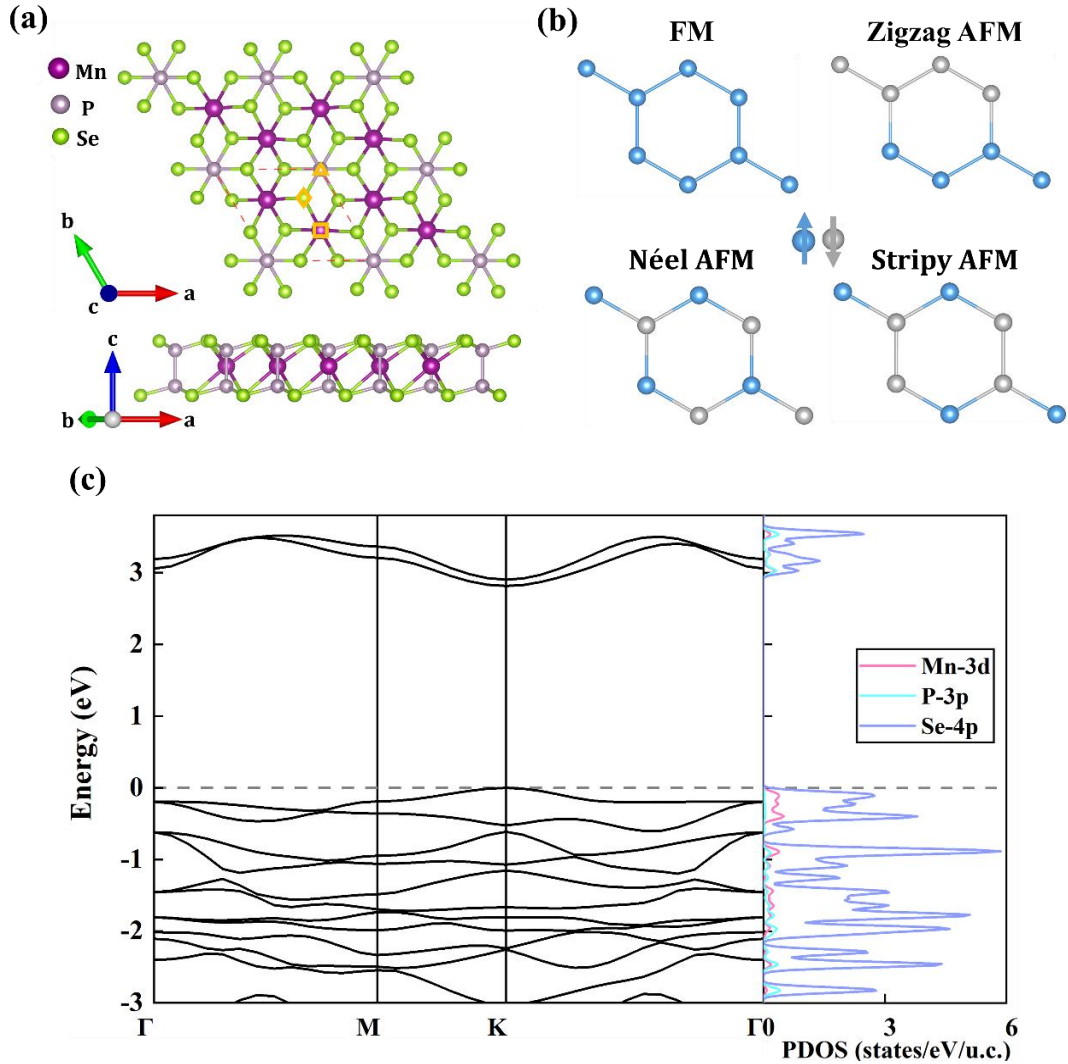

Fig. 1 (a) Top and side views of the monolayer $MnPSe_3$, and the area enclosed by the red dashed lines denotes the unit cell, where the square, rhombus, and triangle denote the three considered adsorption sites; (b) four different magnetic configurations of the FM, zigzag AFM, Néel-AFM and stripy AFM. (c) Band structure and PDOS of monolayer $MnPSe_3$ without SOC, obtained with HSE06 functional. The blue and grey balls are the spin-up and spin-down state Mn atoms, respectively.

The well-documented bandgap underestimation inherent to the PBE functional is evident in our calculations, yielding a HSE06 bandgap of 2.80 eV for monolayer $MnPSe_3$ which slightly exceeding the previous reported value of 2.56 eV [38], as shown in Fig. 1c. This discrepancy persists even with the inclusion of a Hubbard-$U$ correction ($U$ = 4 eV), where PBE+U produces a 1.76 eV gap (Fig. S1), confirming systematic underestimation across exchange-correlation treatments.

Projected density of states (PDOS) analysis reveals that electronic states near $E_F$ are dominated by Se-4$p$ orbitals, with minimal P-3$p$ contributions, consistent with the charge-transfer character of the gap.

## B. Thermodynamic stability of $Li_xMnPSe_3$, $MnPSe_3O_x$ and $MnPSe_3F_x$

Subsequently, we systematically investigated stable adsorption sites and magnetic ground states for Li, F and O adatoms. To model four distinct coverage regimes (25%, 33%, 50%, 100%), we considered single adatom adsorption in four supercell configurations: 1 × 1 × 1, 2 × 1 × 1, √3 × √3 × 1, and 2 × 2 × 1, as illustrated in Fig. S2. For $Li_xMnPSe_3$, $MnPSe_3O_x$ and $MnPSe_3F_x$, three adsorption sites were evaluated (Fig. 1a): above Mn (square), Se (rhombus), and P (triangle). For each configuration, multiple magnetic orderings were considered—two for 1 × 1 × 1, three for 2 × 1 × 1 and √3 × √3 × 1, and four for 2 × 2 × 1—encompassing FM and all symmetry-allowed AFM states. Our calculations reveal a universal preference for Li adsorption at Mn sites across all coverages, while O and F adatoms favor P sites.

To verify the feasibility of adsorption, we calculated the adsorption energy for each atom using the following formula:

$$E_{ad} = E_{total} - E_{MnPSe_3} - E_i. \qquad (1)$$

Here, $E_{MnPSe_3}$ denotes the total energy of the pristine monolayer, $E_{total}$ corresponds to the energy of the adsorbate-substrate system, and $E_i$ represents the energy of an isolated $i$ atom in its free state. As shown in Table 1, the adsorption energies of Li, O, and F on the surface of monolayer $MnPSe_3$ for all four coverages are negative, confirming the feasibility of adsorption. Furthermore, to preclude the formation of bulk elemental phases (e.g., Li-metal, $O_2$, $F_2$), we calculated the cohesive energy $E_{coh} = E_{bulk} - E_{single}$, where $E_{bulk}$ and $E_{single}$ denote the energies of bulk elements and isolated atoms, respectively. The derived values ($E_{coh}$ (Li) = −1.74 eV, $E_{coh}$ (O) = −3.04 eV, $E_{coh}$ (F) = −1.39 eV) are less negative than the corresponding adsorption energies across all coverage regimes (Table 1). This confirms that Li, O, and F adatoms preferentially stabilize as surface-bound species rather than aggregating into bulk phases under the studied conditions.

Table 1 Adsorption energies $E_{ad}$ and $E_{coh(Li)}$ for Li, O, and F adatoms on monolayer MnPSe$_3$ across four coverage regimes (25%, 50%, 75%, 100%).

| Coverage | 25% | 33% | 50% | 100% |
|---|---|---|---|---|
| $E_{ad(Li)}$ (eV) | -2.0 | -1.97 | -2.16 | -2.23 |
| $E_{ad(O)}$ (eV) | -3.28 | -3.76 | -3.83 | -3.44 |
| $E_{ad(F)}$ (eV) | -3.21 | -3.24 | -3.28 | -3.32 |

To assess thermodynamic stability, we first established equilibrium chemical potential constraints for Mn, P, and Se during crystal growth as follow:

$$\Delta\mu_{Mn} + \Delta\mu_P + 3\Delta\mu_{Se} = \Delta H_f(MnPSe_3). \tag{2}$$

Where $\Delta H_f(MnPSe_3)$ is the calculated formation energy of MnPSe$_3$. The stable chemical potential range of MnPSe$_3$ are determined by constraining the possible competing phases of Mn, P, and Se. To prevent the formation of binary phases (MnSe, MnSe$_2$, MnP, MnP$_2$, Mn$_3$P and PSe) and elemental species, the following constraints are imposed on the chemical potentials in this work:

$$\Delta\mu_{Mn} \leq 0, \Delta\mu_P \leq 0, \Delta\mu_{Se} \leq 0, \tag{3}$$

$$\Delta\mu_{Mn} + \Delta\mu_{Se} \leq \Delta H_f(MnSe), \tag{4}$$

$$\Delta\mu_{Mn} + 2\Delta\mu_{Se} \leq \Delta H_f(MnSe_2), \tag{5}$$

$$\Delta\mu_{Mn} + \Delta\mu_P \leq \Delta H_f(MnP), \tag{6}$$

$$\Delta\mu_{Mn} + 2\Delta\mu_P \leq \Delta H_f(MnP_2), \tag{7}$$

$$3\Delta\mu_{Mn} + \Delta\mu_P \leq \Delta H_f(Mn_3P). \tag{8}$$

Phase diagram of the Mn-P-Se system under equilibrium conditions were computed by evaluating competing binary phases, yielding permissible chemical potential ranges (blue region in Fig. 2a). Thus we obtain the chemical potential ranges of Mn, P, and Se in MnPSe$_3$ : $-2.49 \leq \Delta\mu_{Mn} \leq -1.74$; $-0.60 \leq \Delta\mu_P \leq 0$; $-0.30 \leq \Delta\mu_{Se} \leq 0$ (eV). Building on these constraints and incorporating electronegativity considerations, we derived phase diagrams for Li-P-Se, Mn-O, and Mn-F systems (Figs. 2b–d), determining the stable chemical potential windows for Li, O, and F adsorption (blue regions).

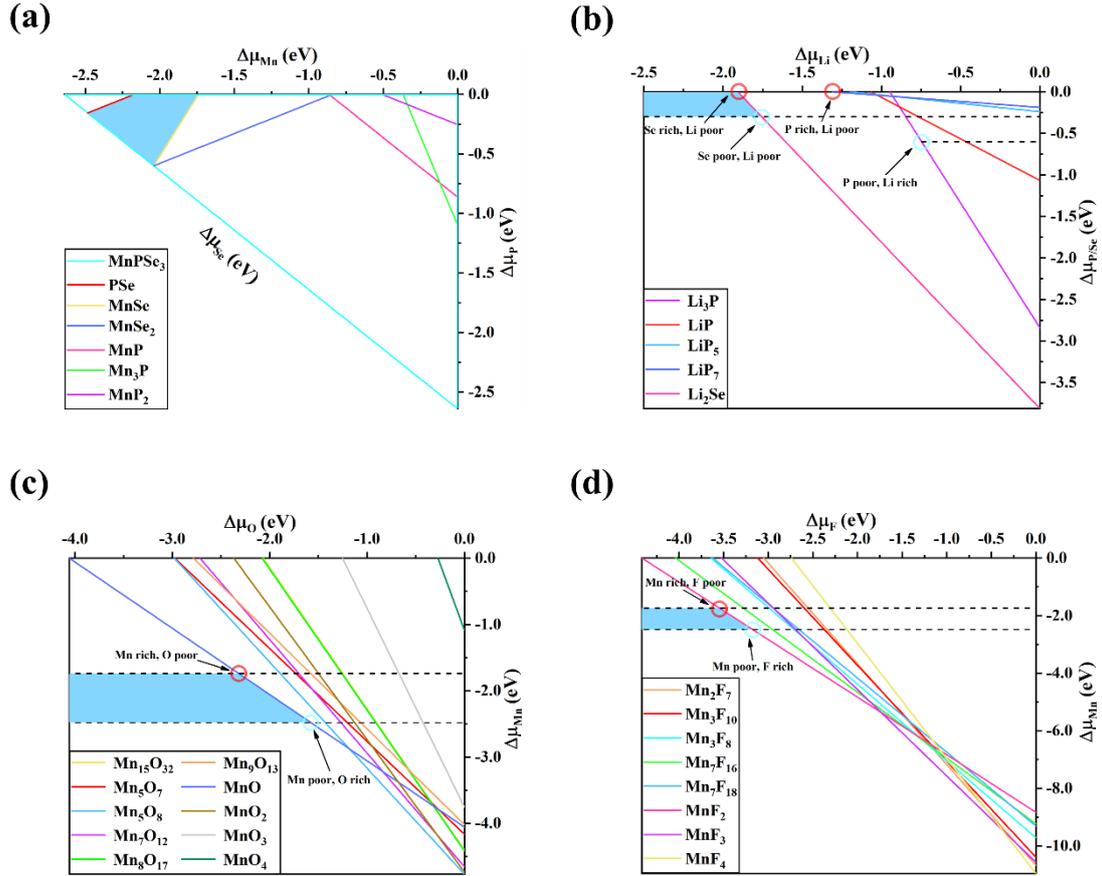

Fig 2. (a) Parametrized phase diagram of $MnPSe_3$ projected onto the ($\Delta\mu_{Mn}$, $\Delta\mu_P$) plane. The blue region denotes the thermodynamically permissible range of Mn, P, and Se chemical potentials under equilibrium conditions; (b) Li-P-Se phase diagram constrained by the permissible $\Delta\mu_P$ and $\Delta\mu_{Se}$ ranges. The blue region defines stable Li chemical potentials ($\Delta\mu_{Li}$); (c) Mn-O phase diagram under the equilibrium $\Delta\mu_{Mn}$ constraint. The blue region indicates stable $\Delta\mu_O$ values; (d) Mn-F phase diagram within the equilibrium $\Delta\mu_{Mn}$ range. The blue region marks permissible $\Delta\mu_F$ values.

As the adsorption energy $E_{ad}$ is consistent with the chemical potential $\Delta\mu$ for the adsorbate, the comparison of coverage-dependent chemical potentials (Table 1) with these stability confinement demonstrates that $Li_xMnPSe_3$ and $MnPSe_3O_x$ remain stable across all four coverages. However, $MnPSe_3F_x$ configurations exceed the maximum F chemical potential in Mn-rich/F-poor conditions (Fig. 2c), achieving stability only when $\Delta\mu_{Mn} < -2.20$ eV. Notably, increasing Li and F adsorption coverage induces progressive shortening of adatom-surface bond lengths, indicative of enhanced interfacial bonding at higher adsorption densities. In contrast, O adsorption exhibits minimal bond-length variation across coverage regimes. This behavior partially explains the anomalous trend observed in Table 1, where higher coverages of Li and F correspond to more negative adsorption energies. Moreover, the low-coverage adsorption configurations of Li, O, and F are intrinsically unstable and tend to coalesce into locally dense regions. Consequently, we restrict our subsequent analyses of magnetic and electronic properties to full-coverage systems ($Li_{0.5}MnPSe_3$, $O_{0.5}MnPSe_3$, and $F_{0.5}MnPSe_3$).

**C. Electronic properties of $Li_{0.5}MnPSe_3$, $MnPSe_3O_{0.5}$ and $MnPSe_3F_{0.5}$**

To investigate the electronic structures of $Li_{0.5}MnPSe_3$, $O_{0.5}MnPSe_3$, and $F_{0.5}MnPSe_3$, we calculated the corresponding band structures (Fig. 3d-f) using the HSE06 method. As shown in Fig. 3(d) and (e), upon adsorption of Li and O atoms, the monolayer $MnPSe_3$ transitions from an AFM semiconductor to an FM semiconductor, with the corresponding band gaps decreasing by approximately 1.9 eV and 2.7 eV, respectively. In contrast, the adsorption of F atom transforms the system into a FM half-metal, as shown in Fig. 3f.

Additionally, we calculated the band structures of $Li_{0.5}MnPSe_3$ and $O_{0.5}MnPSe_3$ using PBE+U (U = 4 eV) method for comparison (see Fig. S3). The resulting band structures indicate that both $Li_{0.5}MnPSe_3$ and $O_{0.5}MnPSe_3$ exhibit half-metallic behavior, further confirming the underestimation of band gaps by the PBE functional.

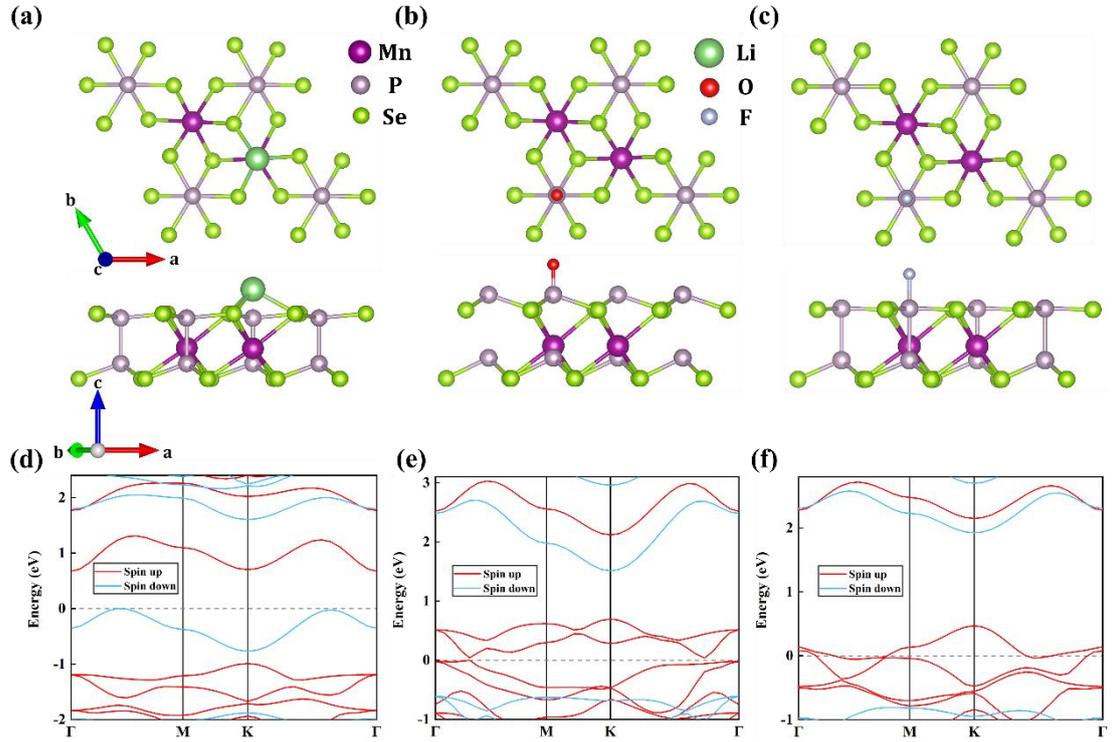

Fig. 3 The most energetically favorable configurations with coverage of 100% after adsorption with (a) Li atom, (b) O atom, and (c) F atom, along with the corresponding band structures for (d) $Li_{0.5}MnPSe_3$, (e) $O_{0.5}MnPSe_3$ and (f) $F_{0.5}MnPSe_3$. The red and blue solid lines represent the spin-up and spin-down channels, respectively. All the shown band structures are obtained with the HSE06 functional.

In addition, we observe Li-adsorption-induced valley polarization in the conduction bands at the K and K' points of monolayer $MnPSe_3$, exhibiting a splitting energy of 20.3 meV as shown in Fig. 4. The pristine monolayer $MnPSe_3$ possesses an in-plane magnetic moment orientation along the y-direction. When SOC is considered, our calculated band structure (Fig. S4) reveals that the valley degeneracy at K and K' points is protected by the combined *MT* symmetry (where *M* represents mirror symmetry and *T* denotes time-reversal symmetry), consistent with the symmetry analysis [39]. The energetically favorable adsorption site for Li atoms is identified at the Mn position, which breaks all mirror symmetry operations. Crucially, our first-principles calculations demonstrate that Li adsorption induces a magnetic easy-axis reorientation from in-plane to out-of-plane direction (detailed discussion in Section D). This structural modification reduces the magnetic point group symmetry from 2/m' in the pristine system to 3.1 in the Li-adsorbed system, consequently breaking the *MT* symmetry protection. The symmetry-breaking mechanism thereby lifts the valley degeneracy and generates measurable valley polarization under SOC conditions.

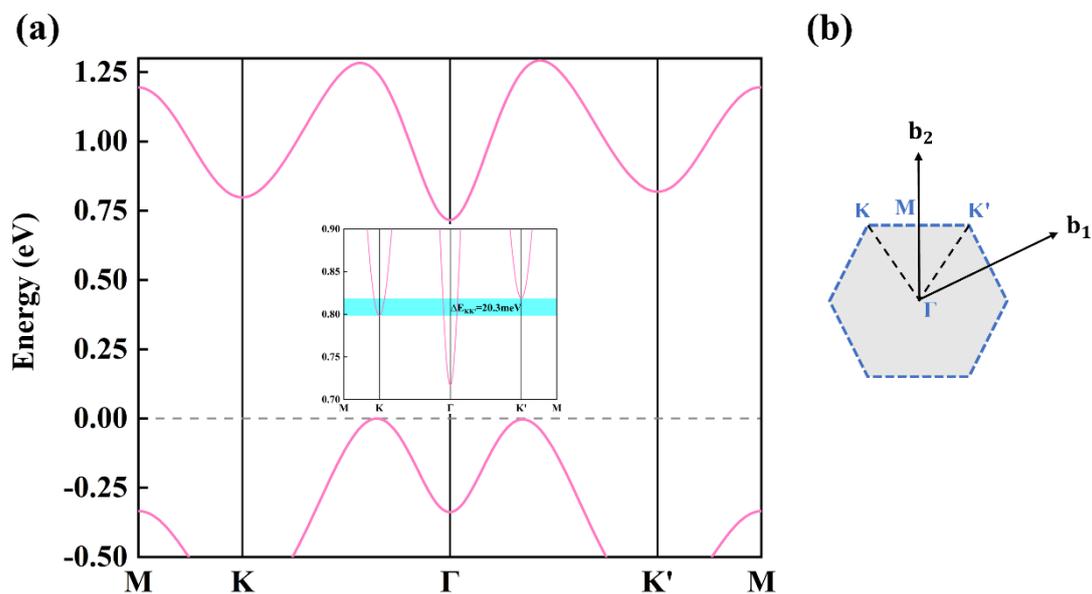

Fig. 4. (a) Band structure of $Li_{0.5}MnPSe_3$ calculated using the HSE06 functional with SOC; (b) Corresponding high-symmetry path in the Brillouin zone.

The crystal structure after adsorption with differential charge densities are presented in Fig. 5, and the electron transfer before and after adsorption are further analyzed using Bader charge calculations. As shown in Fig. 5a, the blue region on Li indicates electron depletion, while the yellow region on Se represents electron accumulation. One can see that charges transfer from Li to the surrounding three Se atoms of monolayer $MnPSe_3$ by -0.31 e/f.u. for Li adsorption, resulting in a reduction of the band gap and $MnPSe_3$ gains electrons. For O and F adsorption, the Bader charge transfers from $MnPSe_3$ to F and O, with transfer amounts of 0.62 and 0.37 e/f.u., respectively, causing the Fermi level to shift downward into $MnPSe_3$, corresponding to hole doping, as shown in Fig 5(b) and (c).

Table 2 The energy difference $\Delta E = E(\text{AFM}) - E(\text{FM})$ in meV, and MAE = $E[001] - E[010]$. The Bader charge difference (e/f.u.) is between before and after Li, O and F adsorption with coverage of 1ML.

| Materials | $MnPSe_3$ | $Li_{0.5}MnPSe_3$ | $MnPSe_3O_{0.5}$ | $MnPSe_3F_{0.5}$ |
|---|---|---|---|---|
| Bader charge (e/f.u.) |  | -0.31 | 0.62 | 0.37 |
| MAE (meV/f.u.) | 0.24 | -0.23 | 0.49 | 1.12 |
| $\Delta E$ (meV/f.u.) | -21.9 | 18.4 | 76.7 | 64.2 |

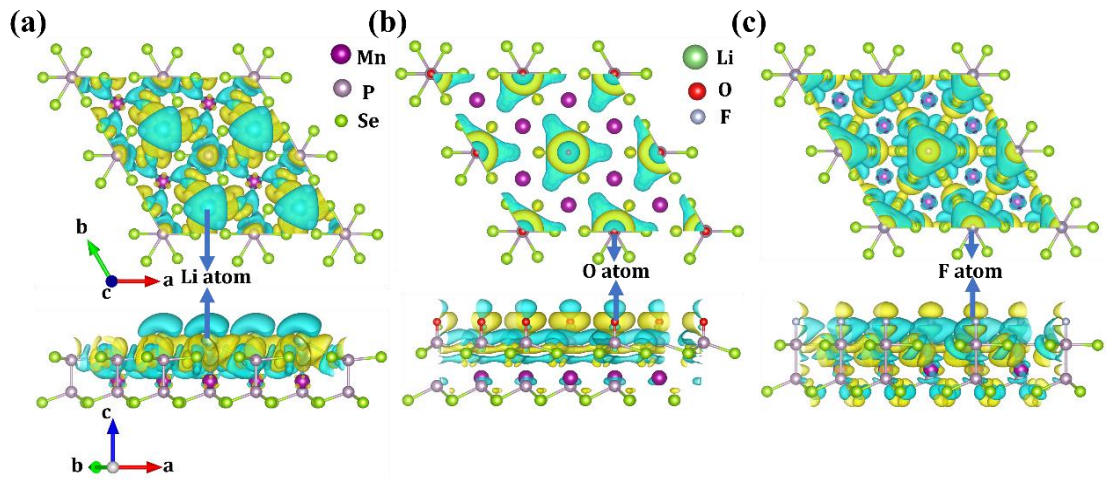

Fig. 5 The crystal structure of (a) $Li_{0.5}MnPSe_3$, (b) $O_{0.5}MnPSe_3$ and (c) $F_{0.5}MnPSe_3$ with different charge densities (isovalue 0.001 e × $A^{-3}$). The blue and yellow distrubutions correspond to the charge depletion and accumulation, respectively

## D. Adsorbate-induced Magnetic Modulation

Next, we calculate the magnetic anisotropy energy (MAE) including SOC to analyze the changes of the easy magnetization axis before and after adsorption. MAE is defined as the total energy difference between $E[001]$ and $E[010]$, i.e., MAE = $E[001] - E[010]$. A positive MAE corresponds to an in-plane magnetization easy axis, while a negative MAE indicates an out-of-plane magnetization easy axis. Our calculations indicate that the pristine monolayer $MnPSe_3$ exhibits an in-plane magnetic easy axis along the y-direction. As shown in Table 2, after Li adsorption, the easy axis of the entire system shifts from in-plane to out-of-plane, while the adsorption of F and O further increases the magnitude of the MAE.

To describe the 2D magnetic structure, the exchange interaction parameters between $Mn^{2+}$ ions spins was extracted by considering corresponding anisotropic Heisenberg Hamiltonian:

$$H = \sum_{i,j} J_1 \vec{S}_i \vec{S}_j + \sum_{i,j} J_2 \vec{S}_i \vec{S}_j + \sum_{i,j} J_3 \vec{S}_i \vec{S}_j + A \sum_i (S_i^z)^2. \tag{9}$$

where $S_i$ represents the total spin magnetic moment of the atomic site $i$. $J_1$, $J_2$, and $J_3$ are the exchange interactions between the first, second, and third nearest-neighbor spins, respectively, as shown in Fig. 6a. A is the single-ion magnetic anisotropic energy, and $S_i^z$ is the spin component of the $i$ site along the z direction of the easy magnetization axis. In the pristine monolayer $MnPSe_3$, all the calculated $J$ values are negative, i.e., $J_1 = -0.45$ meV, $J_2 = -0.04$ meV, and $J_3 = -0.24$ meV, respectively, which show an excellent agreement to $J_1 = -0.46$ meV, $J_2 = -0.03$ meV and $J_3 = -0.19$ meV obtained by linear spin wave theory combined with inelastic neutron scattering which again confirm the robust AFM ground state [40]. As can be seen in Fig. 6c, the Néel temperature of the pristine 2D $MnPSe_3$ is approximately 90 K, which aligns well with previous studies (88 K), thereby validating the reliability of Eq. (9) [10].

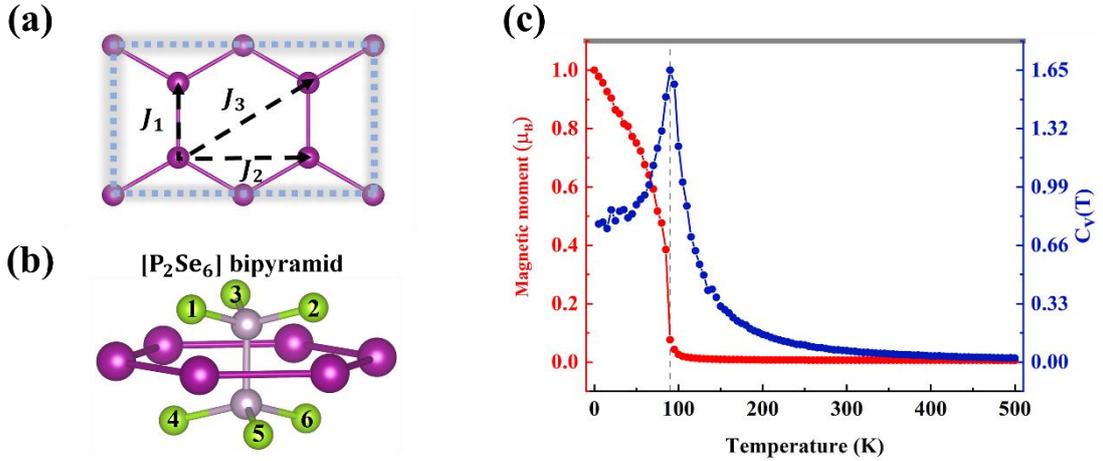

Fig. 6 (a) Magnetic exchange parameters $J_1$, $J_2$, and $J_3$ between $Mn^{2+}$ ions, and (b) Schematic of the Mn-Se⋯Se-Mn super-superexchange pathways governing $J_2$, and $J_3$; (c) The simulated magnetic moment $\mu_B$ and specific heat $C_v$ with respect to temperature for pristine $MnPSe_3$ monolayer.

The magnetic ground state of monolayer $MnPSe_3$ is governed by competing exchange interactions, with first- ($J_1$) and third-nearest-neighbor ($J_3$) couplings dominating. $J_1$ arises from the competition between direct exchange and superexchange interactions. The short-range Mn-Mn direct exchange originates from transitions between Mn ions, where the overlapping half-filled 3$d$-orbitals provide a stable AFM state. Following Goodenough–Kanamori–Anderson (GKA) rules, the near-84° Mn-Se-Mn bond angle favors FM superexchange [41]. However, due to the high-spin state of the Mn ions in the monolayer MnPSe$_3$, the direct AFM interaction dominates over the FM superexchange. For $J_2$ and $J_3$, the AFM coupling originates from different super-superexchange pathways mediated by Se atoms. Specifically, $J_2$ involves the Mn-Se$_3$…Se$_6$-Mn channel, which connects Se atoms from different sublayers, while $J_3$ is associated with the Mn-Se$_1$…Se$_2$-Mn interaction mediated by Se atoms within the same sublayer, as shown in Fig. 6(b) [42]. It is reasonable to expect that the hybridization between the Mn-d states and the Se-p states within the same sublayer is significantly stronger than that between atoms in different sublayers. This explains why $J_3$ is greater than $J_2$, despite the longer Mn–Mn distance in the former. However, $J_1$ is nearly twice as large as $J_3$, thereby ensuring the AFM ordering in monolayer MnPSe$_3$.

Adsorption of Li, O, and F atoms significantly modulates exchange parameters (see Table S1). Li adsorption reduces the bandgap, enhancing Mn-Se-Mn superexchange while increasing Mn-Mn distances to suppress direct exchange. This reverses the $J_1$ dominance, driving an AFM-FM transition. Similarly, O adsorption amplifies next-nearest-neighbor interactions via strengthened $d$-$p$-$d$ orbital hybridization, inducing a more pronounced exchange parameter shift than Li.

The F atom adsorption-induced AFM-FM transition can be qualitatively explained by the Stoner model. According to the Stoner criterion, ferromagnetism emerges when $D(E_F)J > 1$, where $D(E_F)$ is the density of states at the Fermi level in the nonmagnetic state and $J$ represents the strength of the exchange interaction [43]. The exchange parameter $J$ can be estimated by the ratio of the spin splitting of electronic states near the Fermi level to the resulting magnetic moment. Specifically, for the F$_{0.5}$MnPSe$_3$ unit cell, a magnetic moment of 9.2 $\mu_B$ is obtained. As illustrated in the Supplementary Material, the density of states, $D(E_F)$ is 27.8 states/eV, and the average spin splitting between the first two conduction bands and the first valence band is 1.25 eV. There, the calculation result confirms that F$_{0.5}$MnPSe$_3$ satisfies the Stoner criterion ($D(E_F)J > 1$), providing strong evidence that the metallic system exhibits Stoner ferromagnetism.

## E. MAE modulation under carriers doping

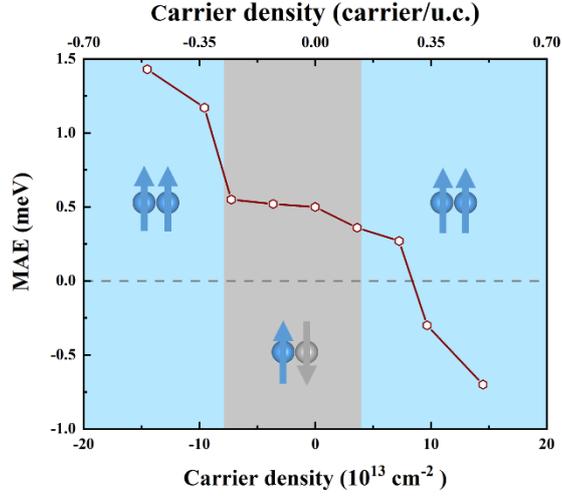

Fig. 7 The variation of MAE (MAE = E[001] - E[010]) with carrier doping concentration. The blue region represents the FM ground state, while the gray region corresponds to the AFM ground state.

Subsequently, carrier doping is applied to modulate the electronic and magnetic properties of monolayer $MnPSe_3$. Fig. 7 displays the variation of MAE under carrier doping. A clear trend is observed, where an increase in hole doping concentration leads to an enhancement of the magnetic anisotropy energy (MAE), stabilizing the in-plane easy axis, while an increase in electron doping concentration continuously decreases the MAE. As the doping concentration reaches a critical threshold, the easy axis flips from in-plane to out-of-plane. It is clear that the Bader charge induced by the Li adsorption (corresponding to electron doping) and O/F adsorption (corresponding to hole doping) exceeds the electron and hole doping mentioned above. Moreover, through comprehensive consideration of multiple antiferromagnetic configurations (Néel-AFM, zigzag-AFM, stripy-AFM), we serendipitously discovered that the phase transition critical concentration ($8 \times 10^{13}$ $cm^{-2}$ for holes and $4.5 \times 10^{13}$ $cm^{-2}$ for electrons) in monolayer $MnPSe_3$ actually surpasses the value reported in previous study [10]. This discrepancy arises from the competitive energy stabilization mechanism, where alternative antiferromagnetic configurations (stripy-AFM and zigzag-AFM) may preferentially occupy lower-energy states under specific doping conditions.

## IV. CONCLUSION

In summary, we have established atomic adsorption as a transformative strategy for tailoring magnetic order and electronic states in monolayer MnPSe$_3$, addressing critical challenges in 2D spintronic material design. Adsorption energy calculations for Li, O, and F at various coverages indicate that the low-coverage adsorption configurations are intrinsically unstable. Instead, these species exhibit a spontaneous tendency to aggregate, leading to the formation of regions with locally high adsorption coverage. Phase diagrams under equilibrium conditions reveal enhanced thermodynamic stability at high coverages. Furthermore, HSE06 calculations demonstrate that full-coverage Li/O adsorption reduces the bandgap, amplifying *d-p-d* superexchange interactions to drive AFM-FM transitions, while F adsorption induces a Stoner ferromagnetism via half-metallic band structure, evidenced by the Stoner criterion ($D(E_\text{F})J > 1$). Li adsorption not only reorients the magnetic easy axis of monolayer MnPSe$_3$ from in-plane to out-of-plane but also induces valley polarization (~20.3 meV) at the K/K' points of the Brillouin zone, demonstrating concurrent control over spin and valley degrees of freedom. Crucially, Carrier doping results indicate that when the electron doping concentration reaches $8.3 \times 10^{13}$ cm$^{-2}$, the magnetic easy axis of monolayer MnPSe$_3$ reorients from an in-plane to an out-of-plane configuration, while the magnitude of the in-plane MAE increases with higher hole doping levels. These insights redefine design principles for two-dimensional magnetic semiconductors by emphasizing the synergy between thermodynamics and spin dynamics. The coverage–stability–magnetism triad delineated herein provides a universal framework for engineering interfacial phenomena in 2D adsorption system, thereby advancing their integration into nonvolatile spintronic memory and logic architectures.

## ACKNOWLEDGMENTS

This work was supported by the Guangdong Basic and Applied Basic Research Foundation (No. 2023A1515012289), and National Natural Science Foundation of China (Grant No. 12474229). This work is partially supported by High Performance Computing Platform of South China University of Technology.

[1] K. S. Novoselov, A. K. Geim, S. V. Morozov, D. Jiang, Y. Zhang, S. V. Dubonos, I. V. Grigorieva, and A. A. Firsov, Electric field effect in atomically thin carbon films, Science **306**, 666

(2004).

[2] A. K. Geim and K. S. Novoselov, The rise of graphene, Nature Materials **6**, 183 (2007).

[3] A. H. Castro Neto, F. Guinea, N. M. R. Peres, K. S. Novoselov, and A. K. Geim, The electronic properties of graphene, Rev. Mod. Phys. **81**, 109 (2009).

[4] N. D. Mermin and H. Wagner, Absence of Ferromagnetism or Antiferromagnetism in One- or Two-Dimensional Isotropic Heisenberg Models, Phys. Rev. Lett. **17**, 1133 (1966).

[5] C. Gong et al., Discovery of intrinsic ferromagnetism in two-dimensional van der Waals crystals, Nature **546**, 265 (2017).

[6] B. Huang et al., Layer-dependent ferromagnetism in a van der Waals crystal down to the monolayer limit, Nature **546**, 270 (2017).

[7] Y. Deng et al., Gate-tunable room-temperature ferromagnetism in two-dimensional $Fe_3GeTe_2$, Nature **563**, 94 (2018).

[8] N. Mounet et al., Two-dimensional materials from high-throughput computational exfoliation of experimentally known compounds, Nature Nanotechnology **13**, 246 (2018).

[9] R. A. de Groot, F. M. Mueller, P. G. van Engen, and K. H. J. Buschow, New Class of Materials: Half-Metallic Ferromagnets, Phys. Rev. Lett. **50**, 2024 (1983).

[10] X. Li, X. Wu, and J. Yang, Half-Metallicity in $MnPSe_3$ Exfoliated Nanosheet with Carrier Doping, J. Am. Chem. Soc. **136**, 11065 (2014).

[11] X. Lin, W. Yang, K. L. Wang, and W. Zhao, Two-dimensional spintronics for low-power electronics, Nature Electronics **2**, 274 (2019).

[12] J. F. Sierra, J. Fabian, R. K. Kawakami, S. Roche, and S. O. Valenzuela, Van der Waals heterostructures for spintronics and opto-spintronics, Nature Nanotechnology **16**, 856 (2021).

[13] S. Yu, J. Tang, Y. Wang, F. Xu, X. Li, and X. Wang, Recent advances in two-dimensional ferromagnetism: strain-, doping-, structural- and electric field-engineering toward spintronic applications, Science and Technology of Advanced Materials **23**, 140 (2022).

[14] J.-J. Xian et al., Spin mapping of intralayer antiferromagnetism and field-induced spin reorientation in monolayer $CrTe_2$, Nature Communications **13**, 257 (2022).

[15] P. Gao, X. Li, and J. Yang, Thickness Dependent Magnetic Transition in Few Layer 1T Phase $CrTe_2$, J. Phys. Chem. Lett. **12**, 6847 (2021).

[16] N. Wang, J. Chen, Y. An, Q. Zhan, and S.-J. Gong, Controllable half-metallicity in $MnPX_3$ monolayer, Npj Spintronics **2**, 60 (2024).

[17] Y. Zhang, Y. Zhang, H. Ye, J. Zhang, and J. Wang, Transition from antiferromagnetic metal to room-temperature ferromagnetic semiconductor in monolayer $CrTe_2$ via Li adsorption, Phys. Rev. B **109**, 195426 (2024).

[18] Y. Gao, X. Jiang, Z. Qiu, and J. Zhao, Photoexcitation induced magnetic phase transition and spin dynamics in antiferromagnetic $MnPS_3$ monolayer, Npj Computational Materials **9**, 107 (2023).

[19] K. Wang, K. Ren, Y. Hou, Y. Cheng, and G. Zhang, Physical insights into enhancing magnetic stability of 2D magnets, Journal of Applied Physics **133**, 110902 (2023).

[20] A. Wiedenmann, J. Rossat-Mignod, A. Louisy, R. Brec, and J. Rouxel, Neutron diffraction study of the layered compounds $MnPSe_3$ and $FePSe_3$, Solid State Communications **40**, 1067 (1981).

[21] D. J. Gillard, D. Wolverson, O. M. Hutchings, and A. I. Tartakovskii, Spin-order-dependent magneto-elastic coupling in two dimensional antiferromagnetic $MnPSe_3$ observed through

Raman spectroscopy, Npj 2D Materials and Applications **8**, 6 (2024).

[22] I. Mazin, R. González-Hernández and L. Šmejkal. Induced Monolayer Altermagnetism in $MnP(S, Se)_3$ and FeSe, arXiv:2309.02355 (2023).

[23] B.-J. Wang, Y.-Y. Sun, J. Chen, W. Ju, Y.-P. An, and S.-J. Gong, Valley splitting in the antiferromagnetic heterostructure $MnPSe_3/WSe_2$, J. Mater. Chem. C **9**, 3562 (2021).

[24] X. Dong, K. Jia, W. Ji, S. Li, and C.-W. Zhang, Realization of Dual Anomalous Valley Hall Effect in Antiferromagnetic $HfN_2/MnPSe_3$ Heterostructure, ACS Appl. Electron. Mater. **6**, 679 (2024).

[25] J. P. Perdew, K. Burke, and M. Ernzerhof, Generalized Gradient Approximation Made Simple, Phys. Rev. Lett. **77**, 3865 (1996).

[26] G. Kresse and J. Furthmüller, Efficient iterative schemes for ab initio total-energy calculations using a plane-wave basis set, Phys. Rev. B **54**, 11169 (1996).

[27] J. Heyd, G. E. Scuseria, and M. Ernzerhof, Hybrid functionals based on a screened Coulomb potential, The Journal of Chemical Physics **118**, 8207 (2003).

[28] X. Wang, D. Wang, R. Wu, and A. J. Freeman, Validity of the force theorem for magnetocrystalline anisotropy, Journal of Magnetism and Magnetic Materials **159**, 337 (1996).

[29] N. Sivadas, M. W. Daniels, R. H. Swendsen, S. Okamoto, and D. Xiao, Magnetic ground state of semiconducting transition-metal trichalcogenide monolayers, Phys. Rev. B **91**, 235425 (2015).

[30] J. Li, Y. Li, S. Du, Z. Wang, B.-L. Gu, S.-C. Zhang, K. He, W. Duan, and Y. Xu, Intrinsic magnetic topological insulators in van der Waals layered $MnBi_2Te_4$-family materials, Sci. Adv. **5**, eaaw5685 (2019).

[31] Y. An et al., Nanodevices engineering and spin transport properties of $MnBi_2Te_4$ monolayer, Npj Computational Materials **7**, 45 (2021).

[32] H. Wu, T. Burnus, Z. Hu, C. Martin, A. Maignan, J. C. Cezar, A. Tanaka, N. B. Brookes, D. I. Khomskii, and L. H. Tjeng, Ising Magnetism and Ferroelectricity in $Ca_3CoMnO_6$, Phys. Rev. Lett. **102**, 026404 (2009).

[33] https://vampire.york.ac.uk/.

[34] R. F. L. Evans, W. J. Fan, P. Chureemart, T. A. Ostler, M. O. A. Ellis, and R. W. Chantrell, Atomistic spin model simulations of magnetic nanomaterials, J. Phys.: Condens. Matter 26, 103202 (2014).

[35] D. J. Gillard, D. Wolverson, O. M. Hutchings, and A. I. Tartakovskii, Spin-order-dependent magneto-elastic coupling in two dimensional antiferromagnetic $MnPSe_3$ observed through Raman spectroscopy, Npj 2D Materials and Applications **8**, 6 (2024).

[36] J. Yang, Y. Zhou, Q. Guo, Y. Dedkov, and E. Voloshina, Electronic, magnetic and optical properties of $MnPX_3$ (X = S, Se) monolayers with and without chalcogen defects: a first-principles study, RSC Adv. **10**, 851 (2020).

[37] B. L. Chittari, Y. Park, D. Lee, M. Han, A. H. MacDonald, E. Hwang, and J. Jung, Electronic and magnetic properties of single-layer $MPX_3$ metal phosphorous trichalcogenides, Phys. Rev. B **94**, 184428 (2016).

[38] K. Du, X. Wang, Y. Liu, P. Hu, M. I. B. Utama, C. K. Gan, Q. Xiong, and C. Kloc, Weak Van der Waals Stacking, Wide-Range Band Gap, and Raman Study on Ultrathin Layers of Metal Phosphorus Trichalcogenides, ACS Nano **10**, 1738 (2016).

[39] Q. Xue, X. Mu, Y. Sun, and J. Zhou, Valley contrasting bulk photovoltaic effect in a *PT*-symmetric $MnPSe_3$ monolayer, Phys. Rev. B **107**, 245404 (2023).

[40] N. Sivadas, M. W. Daniels, R. H. Swendsen, S. Okamoto, and D. Xiao, Magnetic ground state of semiconducting transition-metal trichalcogenide monolayers, Phys. Rev. B **91**, 235425 (2015).

[41] J. Kanamori, Superexchange interaction and symmetry properties of electron orbitals, Journal of Physics and Chemistry of Solids **10**, 87 (1959).

[42] Q. Pei, X.-C. Wang, J.-J. Zou, and W.-B. Mi, Tunable electronic structure and magnetic coupling in strained two-dimensional semiconductor $MnPSe_3$, Frontiers of Physics **13**, 137105 (2018).

[43] E. C. Stoner, Collective electron ferromagnetism, Proceedings of the Royal Society of London. Series A. Mathematical and Physical Sciences **165**, 372 (1938).